\documentclass[%
  aip, 
  jcp, 
  reprint, 
  showpacs,
  amsmath,
  amssymb,
  superscriptaddress,
  floatfix
  ]{revtex4-1}
\usepackage{graphicx}
\usepackage{bm}
\usepackage{color}
\usepackage{ulem}

\newcommand{\rhorOm}{{\rho\left({\mathbf r},{\mathbf \Omega}\right)}}
\newcommand{\rhorom}{{\rho\left({\mathbf r},\Omega\right)}}
\newcommand{\Om}{\mathbf{\Omega}}

\newcommand{\rr}{\mathbf{r}}

\newcommand{\PP}{\mathbf{P}}

\newcommand{\onehalf}{\frac{1}{2}}
\newcommand{\nn}{\nonumber}

\begin{document}

\title{Solvation of Complex Surfaces via Molecular Density Functional Theory}

\author{Maximilien Levesque}
\email{maximilien.levesque@gmail.com}
\affiliation{\'Ecole Normale Sup\'erieure, D\'epartement de Chimie, UMR 8640 CNRS-ENS-UPMC,
24 rue Lhomond, 75005 Paris, France}
\affiliation{CNRS, UPMC Univ. Paris 06, ESPCI, UMR 7195 PECSA, 75005 Paris, France}

\author{Virginie Marry}
\affiliation{CNRS, UPMC Univ. Paris 06, ESPCI, UMR 7195 PECSA, 75005 Paris, France}

\author{Benjamin Rotenberg}
\affiliation{CNRS, UPMC Univ. Paris 06, ESPCI, UMR 7195 PECSA, 75005 Paris, France}

\author{Guillaume Jeanmairet}
\affiliation{\'Ecole Normale Sup\'erieure, D\'epartement de Chimie, UMR 8640 CNRS-ENS-UPMC,
24 rue Lhomond, 75005 Paris, France}

\author{Rodolphe Vuilleumier}
\affiliation{\'Ecole Normale Sup\'erieure, D\'epartement de Chimie, UMR 8640 CNRS-ENS-UPMC,
24 rue Lhomond, 75005 Paris, France}

\author{Daniel Borgis}
\affiliation{\'Ecole Normale Sup\'erieure, D\'epartement de Chimie, UMR 8640 CNRS-ENS-UPMC,
24 rue Lhomond, 75005 Paris, France}

\pacs{%
61.20.Gy, 
61.20.Ja, 
61.25.Em, 
68.03.Hj, 
68.08.-p, 
68.08.De 
}

\date{\today}

\begin{abstract}
We show that classical molecular density functional theory (MDFT), here in the
homogeneous reference fluid approximation in which the functional is inferred from the 
properties of the  bulk solvent,  is a powerful
new tool  to study, at a fully molecular level, the solvation of complex surfaces and interfaces by polar solvents.
This implicit solvent method allows for the determination 
of structural, orientational and energetic solvation properties that are on a par with
all-atom molecular simulations performed for the same system, 
while reducing the computer time by two orders of magnitude.
This is illustrated by the study of an atomistically-resolved
clay surface composed of over a thousand atoms wetted by a molecular dipolar solvent.
The high numerical efficiency
of the method is exploited to carry a systematic analysis of the electrostatic and non-electrostatic
components of the surface-solvent interaction within the popular CLAYFF force field. Solvent energetics and structure
are found to depend weakly upon the atomic charges distribution of the clay surface,
even for a rather polar solvent. We conclude on the consequences of
such findings for force-field development.
\end{abstract}

\maketitle

\section{Introduction\label{sec:Introduction}}

Liquids at solid-liquid interfaces behave differently from bulk liquids.
Interfacial phenomena play key roles in various applications like,
for instance, heterogeneous catalysis, electrochemistry, adsorption
and transport in porous media. It is our goal to improve the description
of such interfaces at the microscopic scale.

Experimentally, specific techniques have been developed, such as the
vibrational sum frequency spectroscopy~\cite{richmond_molecular_2002}
or quasi-elastic neutron scattering techniques~\cite{marry_water_2011},
that now give access to information on the structural and dynamic
properties of the liquid on few molecular layers on top of the surface.
At these scales, theoretical modeling can clarify experimental observations
by linking them to atomistic phenomena. For instance, Rotenberg
\textit{et~al.~}\cite{rotenberg_molecular_2011} showed by molecular
simulations that wetting properties of talc surfaces, 
that behave as hydrophilic or hydrophobic depending on the relative humidity,
are a consequence
of the competition between surface-solvent adhesion, favorable entropy
in the gas phase and molecular cohesion in the liquid phase.

This kind of theoretical modeling is typically based on molecular dynamics (MD) or Monte Carlo simulations
where each atom is considered individually. For systems that are intrinsically
multiscale, e.g. porous media, one has to deal with thousands of atoms.
At this point, numerical heaviness becomes critical and hinders systematic
studies. To overcome this difficulty, various mesoscale models have been proposed,
that are less computationally demanding. One can for instance forget
the molecular nature of the solvent to only account for a Polarizable
Continuum Medium (PCM)~\cite{PCM_leszczynski_computational_1999,PCM_tomasi99,PCM_Tomasi05_review_continuummodels},
or simplify it to a hard sphere, with or without dipole~\cite{DHS_biben_generic_1998,DHS_dzubiella_competition_2004,DHS_oleksy_wetting_2010}.
Coarse-grained strategies have also been successfully applied to solid-liquid
interfaces~\cite{jardat_salt_2009,jardat_self-diffusion_2012}. While
these methods and others proved efficient on a panel of problems,
the applicability of their underlying approximation to an unknown system
remains to be carefully and systematically checked.

In the second half of the last century, liquid state theories~\cite{gray_theory_1984,hansen_theory_2006}
like the classical density functional theory (DFT)~\cite{evans79,evans92,wu_density-functional_2007}
have blossomed. Classical DFT, like other methods such as 
integral equations or classical field theories have shown to give thermodynamic
and structural results comparable to all-atom simulations at attractive
numerical cost, but they have for now been limited to highly symmetric
systems, bulk fluids or simple model interfaces, e.g. hard and soft
walls. A current challenge lies in the development of three-dimensional
theories and implementations to describe molecular liquids, solutions,
mixtures, in complex environments such as biomolecular media or atomically-resolved surfaces and interfaces. 
Recent developments in this direction
use 
lattice field~\cite{lattice_field_azuara_pdb_hydro:_2006,lattice_field_azuara_incorporating_2008},
or Gaussian field~\cite{gaussian_field_varilly_improved_2011} theoretical approaches, or the 3D reference interaction site model (3D-RISM)~\cite{3D_RISM_beglov_integral_1997,3D_RISM_Hirata_Molecular_Theory_of_Solvation,3D_RISM_kloss_quantum_2008,3D_RISM_kloss_treatment_2008,3D_RISM_kovalenko_three-dimensional_1998,3D_RISM_yoshida_molecular_2009}, an appealing integral equation theory that has  proven recently
to be applicable to, e.g.,  structure prediction in complex biomolecular systems. Integral equations are, however,  restricted by the choice
of a closure relation (typically 
HyperNetted Chain (HNC), Percus-Yevick~\cite{percus_analysis_1958}, or Kovalenko-Hirata~\cite{3D_RISM_kovalenko_three-dimensional_1998} approximation), and despite their great potential, they remain
difficult to control and improve, and can prove difficult to converge.

Recently, a molecular density functional theory (MDFT) approach to solvation has been
introduced.~\cite{MDFT_ramirez_density_2002,MDFT_ramirez_density_2005,MDFT_ramirez_direct_2005,MDFT_gendre_classical_2009, MDFT_zhao_molecular_2011, MDFT_borgis_molecular_2012, MDFT_levesque_krfmt} It relies on the definition of a free-energy functional depending on the full six-dimensional position and orientation solvent density. In the so-called homogeneous reference fluid (HRF) approximation, the (unknown) functional can be inferred from the properties of the bulk solvent. Compared to reference molecular dynamics calculations, such approximation was shown to be accurate for polar, non-protic fluids \cite{MDFT_gendre_classical_2009, MDFT_zhao_molecular_2011, MDFT_borgis_molecular_2012}, but to require corrections for water\cite{MDFT_zhao_molecular_2011, MDFT_zhao_new_2011,MDFT_levesque_krfmt,MDFT_zhao_correction_2011}.
Until now, only simple three-dimensional solutes have been considered
in this framework including ions and polar\cite{MDFT_zhao_molecular_2011,MDFT_borgis_molecular_2012} or apolar~\cite{MDFT_levesque_krfmt} molecules. In ref.~\cite{MDFT_borgis_molecular_2012}, the authors have shown that the MDFT scheme was applicable to the computation of electron-transfer properties such as reaction free energies and solvent reorganization free energies.
In this paper, we show how the most recent developments of the molecular density
functional theory, described here in the homogeneous reference fluid approximation~\cite{MDFT_ramirez_density_2002}
(HRF-MDFT), can be implemented to study the solvation of a complex  atomically-resolved clay surface, the pyrophyllite, that contains over a thousand atoms.
In order to uncouple the difficulties intrinsic to
the method to those due to the particular correlations in water, we restrict
ourselves here to a generic model of dipolar solvent, the Stockmayer fluid -- with eventually parameters adjusted to mimick a few properties
of water.

In section \ref{sec:Methods}, we present the most recent developments
in molecular density functional theory in the homogeneous reference
fluid approximation. We also describe our reference molecular dynamics
simulations and the system under investigation. In section~\ref{sec:Results},
we discuss the structural and orientational properties of the complex
interface between the solid and the dipolar liquid. There, we compare
HRF-MDFT with reference molecular dynamics results. Finally, in section~\ref{sec:The-Role-of},
we illustrate the possibilities offered by the method in terms of
numerical efficiency to analyze the electrostatic and non-electrostatic
components of the state-of-the-art force field for clays, CLAYFF~\cite{CLAYFF_cygan_molecular_2004}.
We conclude on the consequences of our analysis on force field development.

\section{Methods and Model System\label{sec:Methods}}

\subsection{Molecular Density Functional Theory of Solvation}

\subsubsection{Theoretical Aspects}

Recent developments of a molecular density functional theory (MDFT) unlocked the
study of the solvation of three-dimensional molecular solutes in arbitrary solvents using classical density functional theory. 
In MDFT, the solvent is composed of rigid molecules and described   in terms of a position and orientation density,
$\rhorom$.  
The solvation free energy is defined as the difference between the grand
potential $\Phi[\rho]$ of a system that includes  external perturbation
(a solute) and inhomogeneous solvent, and the grand potential $\Phi[\rho_{0}]$
of the homogeneous solvent at density $\rho_0 = n_0/8\pi^2$, where $n_0$ is the number density, and without external potential,
\begin{equation}
F[\rho]=\Phi\left[\rho\right]-\Phi[\rho_{0}],\label{eq:solvationF}
\end{equation}
where $\rho$ and $\rho_{0}$ are the position and orientation densities
of the inhomogeneous and homogeneous solvent. 
Following the theoretical framework introduced by Evans~\cite{evans79,evans92,wu_density-functional_2007},
the density functional $F$ can be rewritten as the sum of an ideal
contribution ($F_{id}$), an excess term ($F_{exc}$) and an external
(solute) contribution ($F_{ext}$), 
\begin{equation}
F[\rho] = F_{id}[\rho] + F_{ext}[\rho] + F_{exc}[\rho].
\label{eq:Fsolvdetailed}
\end{equation}
The ideal and external contributions $F_{id}$ and $F_{ext}$ can
be formally and exactly expressed as
\begin{eqnarray}
F_{id}[\rho] & = & k_BT \int d\rr d\Omega [ \rhorom {\rm ln}\left(\frac{\rhorom}{\rho_0}\right ) \\
& & - \rhorom + \rho_0 ] , \label{eq:F_ideal}\\
F_{ext}[\rho] & = & \int d\rr d\Omega\ V_{ext}(\rr,\Omega) \rhorom,
   \label{eq:externo}
\end{eqnarray}
where $k_{B}$ is the Boltzmann constant and $T$ the temperature.
$V_{ext}$ is the sum of the electrostatic and Lennard-Jones interactions
between the solute and one solvent molecule located at $\rr$ with orientation $\Omega$:
\begin{eqnarray}
V_{ext}(\rr,\Omega) &=& \sum_{j \in \text{solvent}}  \{ q_j \, V_q(\rr_j) \nonumber\\
& & + \sum_{i \in \text{solute}}  4\epsilon_{ij}  \left[ \left(\frac{\sigma_{ij}}{r_{ij}} \right)^{12} - \left(\frac{\sigma_{ij}}{r_{ij}} \right)^{6} \right]  \}.
\label{eq:vexteq}
\end{eqnarray}
If $\mathbf{R}(\Omega)$ is the rotation matrix associated to 
$\Omega$ and $\mathbf{s}_j$ is the  position of the solvent site $j$  in the molecular frame, then  $\rr_j = \rr  + \mathbf{R}(\Omega) \mathbf{s}_j$ denotes its absolute position in space and $\rr_{ij} = \rr_j - \rr_i$ its relative position with respect to the solute site $i$ located at $\rr_i$.  $\epsilon_{ij}$, $\sigma_{ij}$ are the Lennard-Jones parameters between solute site $i$ and solvent site $j$.  $q_j$  is the partial charge carried by site $j$ and $V_q(\rr_j)$ is the electrostatic potential created by the solute at $\rr_j$. \\
The excess functional is unknown, but can be expressed formally as 
\begin{eqnarray}
 F_{exc}[\rho] &=& -\onehalf k_BT\iint d\rr_1d\rr_2 d\Omega_1  d\Omega_2 \Delta\rho(\rr_1,\Omega_1) \nonumber\\
 & & \times c(\rr_2 - \rr_1,\Omega_1,\Omega_2) \,  \Delta \rho(\rr_2,\Omega_2) \nonumber\\
 & & + F_B[\Delta \rho],
\label{eq:Fexc}
\end{eqnarray}
where  $\Delta\rhorom =\rhorom - \rho_0$. The first term represents  the homogeneous reference fluid approximation where the excess free-energy density is written in terms of the angular-dependent direct correlation of the {\em pure} solvent. It amounts to a second order Taylor expansion of the excess free-energy functional around the homogeneous solvent. It is equivalent to the HNC approximation in integral  equation theories when the solute is taken as a solvent particle. The second term represents the unknown correction to that term (or bridge term) that can be expressed as of a systematic expansion of the solvent correlations in terms of  the  three-body,\ldots{} n-body terms direct correlation functions. 
We will consider below the case $F_B =0$. This
approximation was shown in Ref.~\cite{MDFT_zhao_molecular_2011} to be accurate for several polar aprotic solvents.
Difficulties were found, however, for water, certainly the most
interesting but also the most complex solvent, in part due to its high-order
correlations, if not interactions. We have proposed several corrections for water entering in the definition of $F_B$: an empirical
three-body correlation term\cite{MDFT_zhao_molecular_2011} inspired by the water model by Molinero \textit{et~al.}~\cite{molinero_water_threebody_2009},
and a bridge function extracted from a hard sphere functional~\cite{MDFT_levesque_krfmt}. The
first one enforces the missing tetrahedral order, while the second one introduces
the $N$-body repulsion terms ($N>2$) of a hard-sphere fluid. \\
Since water adds some extra complexity in the functional that deserve to be tackled separately,  it is not our purpose to study hydration here, but rather  to show the advantages/disadvantages of MDFT and its practical implementation for the molecular solvation of complex 
atomically-resolved surfaces. We will thus consider a generic dipolar fluid, the Stockmayer model, with parameters adjusted to mimick a few properties of water (particle size, molecular dipole and dielectric constant), but no hydrogen bonds, and certainly not a "drinkable" water. \\
For dipolar models, each orientation $\Omega$ can be described by the normalized orientation vector $\Om$ and the direct correlation function in Eq.~\ref{eq:Fexc}
may be developed on a basis of three rotational invariants
\begin{eqnarray}
c\left(\mathbf{r_{12}},\mathbf{\Omega_{1}},\mathbf{\Omega_{2}}\right) &=& c_{S}\left(r_{12}\right)\Phi^{100}\nonumber\\
& &+c_{\Delta}\left(r_{12}\right)\Phi^{110}\left(\mathbf{\Omega_{1}},\mathbf{\Omega_{2}}\right) \nonumber\\
& & +c_{D}\left(r_{12}\right)\Phi^{112}\left(\mathbf{\Omega_{1}},\mathbf{\Omega_{2}}\right),\label{eq:8}
\end{eqnarray}
where 
\begin{eqnarray}
\Phi^{100} & = & 1\label{eq:9},\\
\Phi^{110} & = & \mathbf{\Omega_{1}}\cdot\mathbf{\Omega_{2}},\label{eq:10}\\
\Phi^{112} & = & 3\left(\mathbf{\Omega_{1}}\cdot\mathbf{\hat{r}_{12}}\right)\left(\mathbf{\Omega_{2}}\cdot\mathbf{\hat{r}_{12}}\right)-\mathbf{\Omega_{1}}\cdot\mathbf{\Omega_{2}},\label{eq:11}
\end{eqnarray}
with $\rr_{12} = \rr_{2} - \rr_{1}$ and $\mathbf{\hat{r}_{12}}$ the associated unit vector. \\
If we also define the molecular density $n\left(\mathbf{r}\right)$ and
the polarization density $\mathbf{P}\left(\mathbf{r}\right)$
\begin{eqnarray}
n\left(\mathbf{r}\right) & = & \int\rho\left(\mathbf{r},\mathbf{\Omega}\right)d\mathbf{\mathbf{\Omega}},\label{eq:defn}\\
\mathbf{P}\left(\mathbf{r}\right) & = & \int\mathbf{\mathbf{\Omega}}\cdot\rho\left(\mathbf{r},\mathbf{\Omega}\right)d\mathbf{\mathbf{\Omega}}, \label{eq:p}
\end{eqnarray}
the excess  free-energy density functional $F_{exc}\left[\rho\right]$
can 
be rewritten as a functional of $n$ and $\mathbf{P}$ instead of $\rhorom$\cite{MDFT_ramirez_density_2002,MDFT_ramirez_density_2005}:
\begin{eqnarray}
 F_{exc}\left[ \rhorOm   \right] & = & - \onehalf k_BT \int d\rr_1 d\rr_2  \{ c_S(r_{12}) \, \Delta n(\rr_1)  \, \Delta n(\rr_2) \nn \\
& &-    c_\Delta(r_{12}) \, \PP(\rr_1) \cdot \PP(\rr_1) \\
& &-   c_D(r_{12}) [ 3(\PP(\rr_1) \cdot \hat{\rr}_{12})(\PP(\rr_2) \cdot \hat{\rr}_{12}) \nonumber\\
& & - \PP(\rr_1) \cdot \PP(\rr_1) ]  \}. \nn
\end{eqnarray}
This significantly
increases numerical efficiency, as will be discussed below. The same reduction is true for the ideal and external contributions  if the solute-solvent electrostatic interaction is  strictly restricted to charge-dipole 
interactions\cite{MDFT_ramirez_density_2002,MDFT_ramirez_density_2005}. Even in that case, however, it remains advantageous, both for convergence reasons and for keeping the generality of the code, to stick to the expression of the ideal free-energy in terms of the angular distribution $\rhorOm$, eq.~\ref{eq:F_ideal}, and to minimize the functional in the full position-angle space. This is now described.

\subsubsection{Numerical Aspects of HRF-MDFT}

Here, we give insight into numerical details associated with the variational
minimization of the density functional $F$, \textit{i.e.} the resolution
of the Euler-Lagrange equation
\begin{equation}
\frac{\delta F[\rho]}{\delta\rho}=k_{B}T\ln\left(\frac{\rho\left(\mathbf{r},\mathbf{\Omega}\right)}{\rho_{0}}\right)+V_{ext}\left(\mathbf{r},\mathbf{\Omega}\right)+\frac{\delta F_{exc}[\rho]}{\delta\rho}=0.
\end{equation}
The inhomogeneous density $\rho$ is projected onto an orthorhombic
position grid of $N_{x}\times N_{y}\times N_{z}$ nodes with periodic boundary
conditions. To each node is associated an angular grid on which is
discretized the orientation vector $\mathbf{\Omega}$. The variational
density $\rhorOm$ which minimizes $F$ is optimized
numerically by the limited memory Broyden\textendash{}Fletcher\textendash{}Goldfarb\textendash{}Shanno
(L-BFGS) method as implemented by Byrd, Lu, Nocedal and Zhu~\cite{LBFGS_byrd_limited_1995,LBFGS_zhu_algorithm_1997}.
This quasi-Newton algorithm only requires the knowledge of the functional
$F$ and its first derivative with respect to the density at each
node, which are known analytically. The Hessian, needed in Newton-derived
algorithms, is approximated using gradients at previous self-consistent
iterations. This makes a notable difference with other optimizers:
faster than Piccard iterations, without requiring the second variation
of the functional with respect to the density as typically required
by pseudo-Newton algorithms. Convolutions in $F_{exc}$ are calculated
in reciprocal space by fast Fourier transforms (FFT) as implemented
in the FFTW3 library~\cite{FFTW_frigo_fftw,FFTW_frigo_design_2005,FFTW_url}.
Rewriting $F$ as a functional of $n$ and $\mathbf{P}$ reduces considerably
the number of angular summations and of FFTs to be performed. As a summary, the high performance
of our three-dimensional HRF-MDFT is due to ($i$) a quasi-Newton
implementation, ($ii$) the calculation of convolutions in reciprocal
space, associated to the extreme performance of FFTW3, ($iii$) the
rewriting of the functional in a form optimal for our purpose.

\begin{figure}
\begin{centering}
\includegraphics[width=8.5cm]{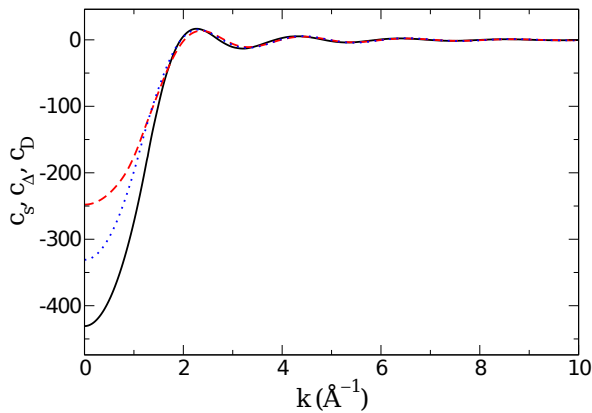}
\par\end{centering}

\caption{\label{fig:Projections}Projections $c_{S}$ (black line), $c_{\Delta}$
(red dashed line) and $c_{D}$ (blue dotted line) of the direct correlation
function of the Stockmayer fluid on the first three rotational invariants. }

\end{figure}

For the given solvent molecular model, the direct correlation function of the solvent is extracted from all-atom
simulations from which are generated the pair distribution functions.
The corresponding correlation function can then be deduced by solving
the Ornstein-Zernike equation. This can be done in Fourier space
but raises numerical issues at large $r$, \textit{i.e.} at small
$k$, the conjugate of $r$. The direct space method of Baxter combined
with the variational method of Dixon and Hutchinson was used accordingly~\cite{dixon_method_1977,MDFT_ramirez_density_2002}.
This method enjoins that $c$ vanishes beyond a radius, set to 8.7~\AA \, in the present system.
The projection of the direct correlation function of the Stockmayer
fluid on rotational invariants $c_{S}$, $c_{\Delta}$ and $c_{D}$
as expressed in Eqs.~\ref{eq:8} to \ref{eq:11} are plotted as a
function of $k$ in figure~\ref{fig:Projections}. Note that the exact bridge
function of the Stockmayer fluid from explicit Monte Carlo simulation
data has been published recently by Puibasset and Belloni~\cite{puibasset_bridge_2012}.

The external potential $V_{exc}(\mathbf{r},\mathbf{\Omega})$ is computed only once, at the beginning of the simulation:
At each node, an isolated solvent molecule is considered with a given molecular orientation, for which is tabulated
the total interaction energy between the solvent molecule sites and the solute sites,
according to Eq.~\ref{eq:vexteq}. To accelarate the computation, the value of the electrostatic potential at each solvent site is interpolated from its values on 
the grid. The grid electrosatic potential itself is obtained by first extrapolating the solute charge density  
on the grid and then solving the resulting Poisson equation by FFT's. The calculation of the Lennard-Jones
external potential takes advantage of cut-off distances for each solute site.

\begin{figure}
\begin{centering}
\includegraphics[width=8.5cm]{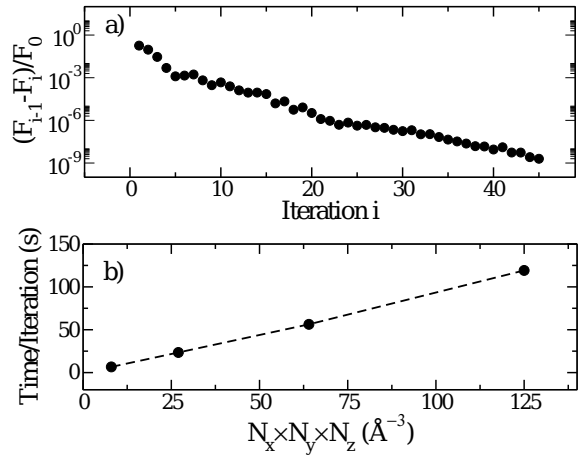}
\par\end{centering}

\caption{\label{fig:Convergence-pour-}a) Variation of the solvation free energy
functional $F$ between two iterations normalized by the initial value
$F_{0}$ during the iteration process. The grid is composed of $4\times4\times4$
nodes per \AA$^{3}$ and 18~discrete orientations per node. A typical
minimization is performed with a precision of $10^{-5}$ in the solvation energy in less than 15 iterations.
b) CPU time in seconds per iteration
for grid meshes $N_{x}\times N_{y}\times N_{z}=2^{3}$, $3^{3}$,
$4^{3}$ and $5^{3}$ per \AA$^{3}$ and 18~discrete orientations
per node. The calculation of the external potential for a given grid mesh is done only once at the beginning of the simulation and takes a CPU time comparable to 1 to 2 iterations. The solvated solute consists of 1280 point charges and 1152~Lennard-Jones
sites, for which parameters are extracted from the popular CLAYFF force field~\cite{CLAYFF_cygan_molecular_2004} and given
in table~\ref{tab:Force-field-used}. The supercell volume is approximately $68$~nm$^{3}$.}

\end{figure}

In this work, we used typically $4\times4\times4$ nodes per \AA$^{3}$, which, for the system considered
below, corresponds to roughly $150^3$ 3D-grid points.
The molecular orientation $\mathbf{\Omega}$ was discretized over 18~angles
and integrated over the whole sphere by Legendre quadratures. This
implies $4\times4\times4\times18=1152$ variables per \AA$^{3}$
to be optimized. All results presented here were carefully checked
with respect to the number of nodes per \AA$^{3}$ and per orientation.
The convergence of the results as a function of grid resolution can be quantified as follows:
 with respect to calculations with a very fine grid resolution of $5^3$ grid points per \AA$^3$,
the relative difference in solvation free energy amounts to 1.2~\% with $2^3$ grid points per \AA$^3$ and $0.3$~\% with  $3^3$ grid points per \AA$^3$.
It is less than 0.1~\% with $4^3$ grid points per \AA$^3$.
Values given above are found to be adequate for all observables
presented in this paper. In order to illustrate the high efficiency
of the method, the iterative convergence of $F$ with iteration steps
is illustrated in figure~\ref{fig:Convergence-pour-} for the grid
described above. After 15~iterations, the convergence in solvation
free energy is of the order of $10^{-5}$. We also illustrate in figure~\ref{fig:Convergence-pour-}
the CPU time per iteration for 18~angles per node and $2^{3}$, $3^{3}$,
$4^{3}$ and $5^{3}$ nodes per \AA$^{3}$ for our model system containing
several hundred atoms. One observes a linear scaling in $N_{x}\times N_{y}\times N_{z}$,
which leads to convergence in less than 15~minutes for the $4\times4\times4$
grid per \AA$^{3}$~described above on an ordinary laptop\footnote{Intel Core~i7 processor at 2.7 GHz with 8~Gigabytes of RAM} without parallelization.\\

\subsection{Molecular Dynamics\label{sub:Molecular-Dynamics}}

Molecular dynamics simulations were generated for the same surface and solvent molecular models in order to
compare with the MDFT results. 
All simulations were done in the canonical NVT ensemble, with a Nose-Hoover
thermostat. After a phase of equilibration, all structural quantities
were collected and averaged over 5~ns. The local molecular density
$n(\mathbf{r})$ of the Stockmayer fluid is calculated as the time
averaged density in elementary volumes of $0.1^{3}$~\AA$^3$.
The densities perpendicularly to the clay layers can be calculated
by averaging the above density in the $(x,y)$ planes. The spatial
densities of the average dipoles were calculated in the same way. Molecular
dynamics simulations were performed with the DLPOLY package~\cite{DL_POLY_todorov_dl_poly_3,DL_POLY_url}.

\subsection{System Description\label{sub:System-Description}}

Pyrophyllite is neutral clay. It is monoclinic of space
group $2/m$, perfectly cleaved along orientation $\{001\}$. Two
views of a pyrophyllite sheet are provided in figure~\ref{fig:Microscopic-clay-structure,}.
It consists of a stack of 6 atomic layers. The top layer consists
of oxygen (O) and silicon (Si) atoms with a lateral hexagonal symmetry.
The central layer consists of aluminum (Al) atoms in $2/3$ of the octahedral
sites. Between these two layers is a oxygen-hydrogen (O-H) layer,
with O atoms at the center of the hexagons formed by the top Al-O
sites. The O-H axis is oriented in the direction of the empty octahedral
site. The coordinate~$z$ of each site along the normal to the clay surface is given in table~\ref{tab:coordinate}.
The simulation box contains two half clay layers of lateral
dimensions $L_{x}\times L_{y}=41.44\times35.88$~\AA$^{2}$, which
corresponds to 32~clay unit cells of formula Al$_{4}${[}Si$_{8}$O$_{20}${]}(OH)$_{4}$.
The distance between the surfaces is $L_{z}=45.57$~\AA, chosen
to recover the bulk density of the Stockmayer fluid at the center of
the pore,\textit{ i.e.} $n_{0}=0.0289$~molecule per \AA$^{3}$. The
simulation supercell is thus $\approx68$~nm$^{^{3}}$ in periodic
boundary conditions, and contains 1280~clay atoms (640 per half clay
layer). This pore contains 1600~solvent
molecules in the reference all-atom simulations.\\
\begin{figure}
\begin{centering}
\includegraphics[width=8.5cm]{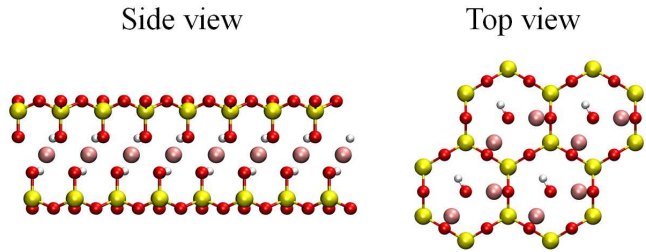}
\par\end{centering}

\caption{\label{fig:Microscopic-clay-structure,}Microscopic clay structure,
side and top views (red, O; white, H; yellow, Si; cyan, Al). In the
top view, only layers above the Al atoms are shown to highlight the
hexagonal symmetry of the first O-Si layer, the hydroxyl group (-OH)
parallel to the sheet, and the Al position in $2/3$ of the octahedral sites.
}
\end{figure}

\begin{table}
\centering{}%
\begin{tabular}{|c|c|c|c|c|c|c|c|c|c|c|}
\cline{1-1} \cline{3-11} 
type &  & O & Si & O & H & Al & H & O & Si & O\tabularnewline
\cline{1-1} \cline{3-11} 
$z$~(\AA) &  & 0.00 & 0.59 & 2.18 & 2.27 & 3.27 & 4.27 & 4.36 & 5.95 & 6.54\tabularnewline
\cline{1-1} \cline{3-11} 
\end{tabular}\caption{\label{tab:coordinate}Coordinate $z$ along the
normal to the surface of each site of the pyrophyllite. The distance
$\Delta z$ between two successive layers is also indicated.}
\end{table}

We used the CLAYFF force field for both molecular dynamics simulations and MDFT
minimization. It is a general purpose force field for simulations involving
multicomponent mineral systems and their interfaces with solutions.
Related information is tabulated in table~\ref{tab:Force-field-used}.
The Stockmayer fluid is described by a 3-sites molecule. The neutral
central site interacts via Lennard-Jones potentials only. The external
sites have opposite charges of $\pm1.91$~e, both distant of 0.1~\AA~from
the central site. It results in a solvent molecular dipole of~1.84~D (incidentally, approximately
that of a water molecule in the gas phase), and,
with the chosen Lennard-Jones parameters that match the reduced parameter set of Pollock and Alder~\cite{pollock_static_1980},
in a dielectric constant of roughly 80, comparable to that of bulk water at room conditions~\cite{MDFT_ramirez_density_2002}. 
The parameters for the Stockmayer fluid force field can also be found
in table~\ref{tab:Force-field-used}. Once again, it is not our purpose
here to study the more complex solvation by water, which requires extra terms in the functional, but to demonstrate the possibilities
of MDFT in the HRF approximation for a generic polar solvent whose functional
is of a good quality. \\
\begin{table}
\begin{centering}
\begin{tabular}{|c|c|c|c|c|}
\hline 
Molecule & Atom & $\epsilon$~(kJ/mol) & $\sigma$~(\AA) & $q$~(e)\tabularnewline
\hline 
\hline 
Pyrophyllite & Al & 5.56388e-6 & 4.27120 & 1.575\tabularnewline
\cline{2-5} 
 & Si & 7.7005e-6 & 3.30203 & 2.100\tabularnewline
\cline{2-5} 
 & O$_{G}$ & 0.650190 & 3.16554 & $-1.050$\tabularnewline
\cline{2-5} 
 & O$_{H}$ & 0.650190 & 3.16554 & $-0.950$\tabularnewline
\cline{2-5} 
 & H$_{G}$ & 0.0 & 0.0 & 0.425\tabularnewline
\hline 
Stockmayer & central & 1.847 & 3.024 & 0.0\tabularnewline
\cline{2-5} 
 & side 1 & 0.0 & 0.0 & 1.91\tabularnewline
\cline{2-5} 
 & side 2 & 0.0 & 0.0 & $-1.91$\tabularnewline
\hline 
\end{tabular}
\par\end{centering}

\caption{\label{tab:Force-field-used}Force field used to model the pyrophyllite
solute and the Stockmayer fluid in both molecular dynamics simulations and HRF-MDFT
minimization. Pyrophyllite parameters are extracted from the CLAYFF
force field~\cite{CLAYFF_cygan_molecular_2004}.}
\end{table}

\section{Results\label{sec:Results}}

We first compare the predictions of HRF-MDFT for the solvent density
and orientation to reference all-atom simulations, which are two orders of magnitude slower to acquire. We then analyze the role of electrostatic interactions
on these quantities.

\subsection{Density Profiles\label{sub:Density-Profiles}}

\begin{figure}
\begin{centering}
\includegraphics[width=8.5cm]{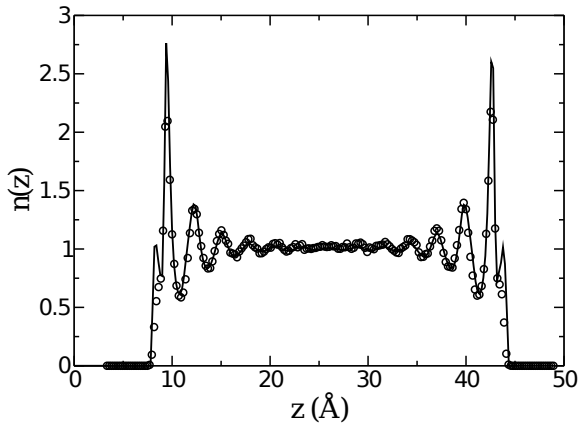}
\par\end{centering}

\caption{\label{fig:Density-profile-along}Normalized number density profile of the solvent
between two pyrophyllite layers as calculated by molecular dynamics
(open circles) and HRF-MDFT (black line).}
\end{figure}

The main structural observable describing solvated interfaces is the
number density profile $n(z)$, defined as the average of the number
density on plane $z$,
\begin{equation}
n(z)=\frac{1}{L_{x}L_{y}}\iint\frac{n(\mathbf{r})}{n_{0}}dxdy.
\end{equation}
The density profiles extracted from explicit molecular dynamics and
from HRF-MDFT are given in figure~\ref{fig:Density-profile-along}.
Good overall agreement is found between the two methods. A shoulder
followed by two main peaks are found in MD. In HRF-MDFT, the shoulder
looks more like a weak peak. The two strongest peaks are followed
by long-ranged oscillations. The first feature (shoulder or weak peak)
is found at $z=8.3$~\AA, 1.76~\AA ~away of the surface layer
(see Table~\ref{tab:coordinate} to identify layers
coordinates). Its intensity is limited to $\approx0.7$ in MD and
$\approx1$ in HRF-MDFT. The largest and main peak is at $z=9.5$~\AA,
2.96~\AA~after the top surface layer. An intermediately intense
and broad peak is also found at $z=12.3$~\AA, 5.76~\AA~away
of the surface. These three structures will later be called ``prepeak'',
``main peak'' and ``secondary peak''. Further weak oscillations
are found up to 15~\AA~away from the surface. At the center of
the supercell, the density is flat at $n=n_{0}$, namely at the bulk
density, which means for solvent molecules to be in bulk, homogeneous,
conditions.\\

\begin{figure}
\begin{centering}
\includegraphics[width=8.5cm]{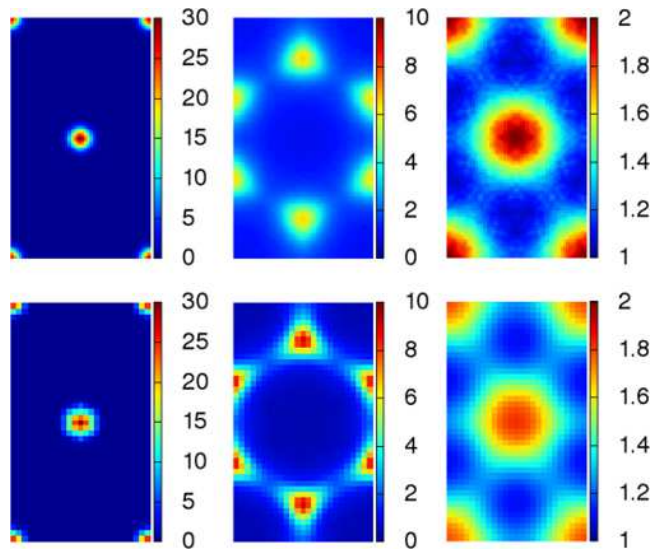}
\par\end{centering}

\caption{\label{fig:Solvant-density-in}Maps of the solvent normalized number density
$n(\mathbf{r})/n_{0}$ in planes of the prepeak (left), main peak
(center) and secondary peak (right), as calculated by molecular dynamics
(top) and HRF-MDFT (bottom).}
\end{figure}

The only noticeable difference lies in the weak prepeak seen in HRF-MDFT
observed as a shoulder in molecular dynamics. The HRF-MDFT is known
to slightly overestimate the height of the first peak in polarized
systems~\cite{MDFT_zhao_molecular_2011}. The localized nature of
the prepeak can be seen in figure~\ref{fig:Solvant-density-in} where
is presented the density map in the plane defined by the maximum of
the prepeak, as calculated by MD and HRF-MDFT. The same position and
overall shape is found with the two methods. It is slightly broader
in-plane in HRF-MDFT for approximately the same intensity, which induces the higher value once averaged
in the plane. The prepeak is localized at the center of the hexagons
formed by surface Si and O atoms. One may note the high maximum value
of the normalized number density $n/n_{0}$ there (up to $\approx30$). The integral
of the density in this peak is the total number of particles in it.
While a strict deconvolution of the three-dimensional prepeak is arbitrary,
no consistent definition results in more than one solvent molecule
inside each prepeak.

In figure~\ref{fig:Solvant-density-in}, we also plot the number
density in the plane of the main and secondary peaks. Again, the shapes
are very similar in MD and HRF-MDFT. The main peak is localized on top
of Si atoms. On the contrary, a depletion is found on top of O atoms
of the surface layer. The broad secondary peak can be found again
on top of the center of hexagons, \textit{i.e.} on top of O atoms.

We now have a clear three-dimensional view of the solvent structure
on the pyrophyllite surface: rare solvent molecules are adsorbed very
close to the surface, at the center of hexagons formed by Si and O
atoms of the clay surface layer. On top of these molecules
is stacked a strongly structured hexagonal layer of solvent molecules,
above Si atoms. An additional, more diffuse layer is found once again
on top of the center of the hexagons. The relatively small distance between the
first two layers demonstrates significant interactions,
and thus cohesion, between layers. This conclusion is in agreement
with Rotenberg \textit{et~al.} who showed recently how the competition
between adhesion and cohesion determines the hydrophobicity of these
surfaces~\cite{rotenberg_molecular_2011}.\\
As a partial conclusion about structural properties, HRF-MDFT results
are in quantitative agreement with the reference molecular dynamics
simulations, with a speed-up of \textit{two orders of magnitude}.

\subsection{Orientational Properties\label{sub:Orientational-Properties}}

We now turn to the orientational properties of the molecules solvating
the pyrophyllite surface. Even if it is routinely done, it is worth
emphasizing that orientational properties are slower to sample in
molecular dynamics or Monte Carlo simulations as each elementary volume
described in section~\ref{sub:Molecular-Dynamics} has to be sampled
for a number of angles. This problem is recurrent in numerous simulation
techniques where orientational degrees of freedom (from electronic spins
to molecular orientations) induces new and rich behaviors~\cite{lavrentiev_magnetic_2010,levesque_simple_2011,levesque_electronic_2012}.
For its part, HRF-MDFT gives a direct access to this quantity through
the full orientational density $\rhorOm$, and also the polarization density $\mathbf{P}(\rr)$ defined in Eq.~\ref{eq:p},
as one of the two natural observables of the theory.

\begin{figure}
\begin{centering}
\includegraphics[width=8.5cm]{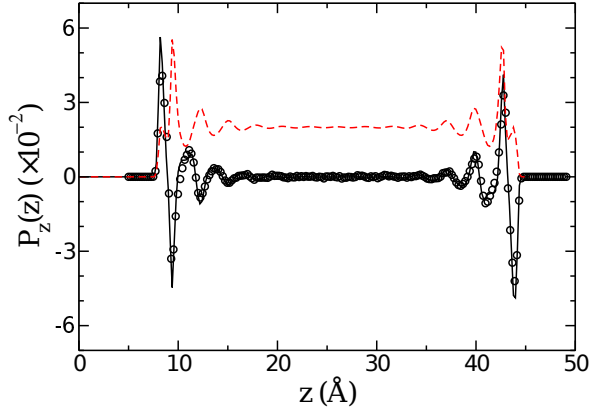}
\par\end{centering}

\caption{Component along the normal $z$ to the surface of the dimensionless
solvent polarization density, between two pyrophyllite layers, as
calculated by molecular dynamics (open circles) and HRF-MDFT (black
line). The density profile is also presented in arbitrary units (red
dashed line). Maxima in the polarization density correspond to maxima
in the solvent number density.\label{fig:pz_of_z_md_mdft}}

\end{figure}

\begin{figure}
\begin{centering}
\includegraphics[width=8.5cm]{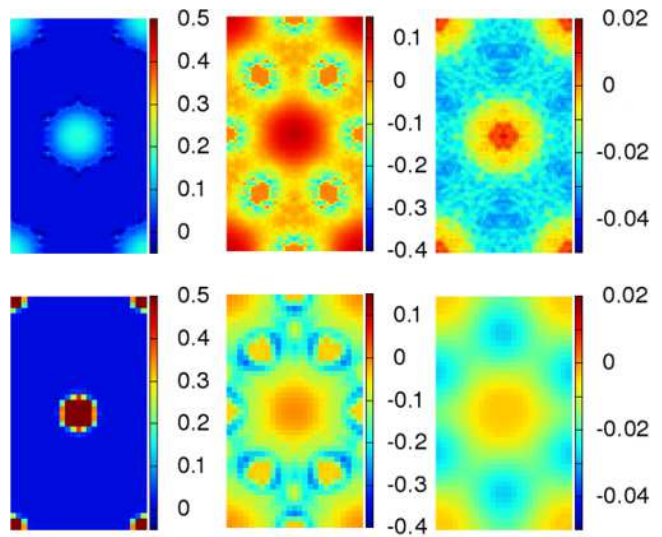}
\par\end{centering}

\caption{Maps of the polarization density projected on the normal $z$ to the
surface axis, $P_{z}$, in the planes of the prepeak (left), main
peak (center) and secondary peak (right), as calculated by molecular
dynamics simulations (top) and HRF-MDFT (bottom).\label{fig:Polarization-density-maps}}

\end{figure}

The polarization density is found to be aligned in the $z$ direction, both in MD and HRF-MDFT. In figure~\ref{fig:pz_of_z_md_mdft},
we report the projection of $\mathbf{P}$ on the $z$ axis (noted
$P_{z}$) as a function of coordinate $z$ (\textit{i.e.} it is averaged
in each plane $z$). Quantitative agreement is again found between
the reference molecular dynamics simulations and the HRF-MDFT. A first
peak is found at the location of the prepeak, another one at the main
peak, \textit{etc}. Between each maximum, the sign of $P_{z}$ changes.
Maps of $P_{z}$ in the planes of the prepeak, main peak and secondary
peak are given in figure~\ref{fig:Polarization-density-maps}. Similarly to the density, the overall shapes from both methods are in very
good agreement. The polarization density is overestimated by HRF-MDFT
in the prepeak, which is expected from previous work on small polarized
systems~\cite{MDFT_zhao_molecular_2011} where the polarization close
to the solute is often slightly overestimated. In the main and secondary peaks,
the polarization densities are found larger in MD than in HRF-MDFT,
although not significantly. This may be due to a balance of the layer-by-layer
polarizations in HRF-MDFT.

In order to get more information on the orientational properties per
molecule, we also define several orientation order parameters. One
is the local orientation of the polarization density, defined as the
cosine of the angle between the polarization density $\mathbf{P}(\mathbf{r})$
and the normal $\mathbf{z}$ to the surface,
\begin{equation}
\cos\theta_{P}(\mathbf{r})=\frac{P_{z}\left(\mathbf{r}\right)}{\left\Vert \mathbf{P}(\mathbf{r})\right\Vert }.\label{eq:costhetap}
\end{equation}
Another is the averaged molecular orientation 
\begin{equation}
\cos\theta_{\mu}(\mathbf{r})=\frac{\left\Vert P_{z} \left(\mathbf{r}\right) \right\Vert}{n\left(\mathbf{r}\right)}.\label{eq:costhetamu}
\end{equation}

\begin{figure}
\begin{centering}
\includegraphics[width=8.5cm]{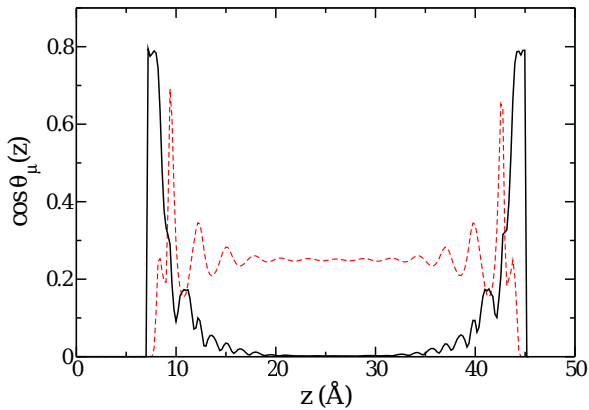}
\par\end{centering}

\caption{\label{fig:Profile-along-the}Profile along the normal to the surface
$z$ of the average molecular orientation,~$\cos\theta_{\mu}(z)$
(black line). The density profile of the solvent is also given in
arbitrary units (red dashed line).}

\end{figure}

We plot the average molecular orientation $\cos\theta_{\mu}(z)=\left\langle \cos\theta_{\mu}(\mathbf{r})\right\rangle _{xy}$
in figure~\ref{fig:Profile-along-the}. The solvent molecules show
strong average orientation in the first layer. On the contrary, the
high polarization density in the second, more cohesive  layer reflects
the large number of molecules inside, each one with a relatively small
preferential orientation along $z$.

\begin{figure}
\begin{centering}
\includegraphics[width=8.5cm]{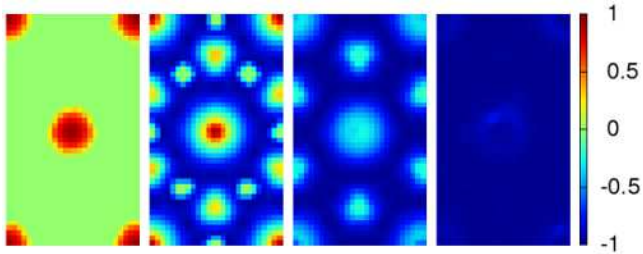}
\par\end{centering}

\caption{\label{fig:The-order-parameter}From left to right, order parameter
$\cos\theta_{P}$ in the prepeak, at the intermediate state between
the prepeak and the main peak, in the main peak, and in the secondary
peak.}
\end{figure}

One can get more insight into the orientation properties of the solvent
molecules by looking at the order parameter $\cos\theta_{P}$ defined
in Eq.~\ref{eq:costhetap}. When $\cos\theta_{P}=1$ ($-1$), the
dipole is oriented against (toward) the surface on the left of the supercell, and vice versa for the other surface. In figure~\ref{fig:The-order-parameter},
we plot $\cos\theta_{P}$ in the plane of the first three peaks identified
earlier and in an intermediate coordinate $z$ between the prepeak
and the main peak. We observe that solvent molecules in the prepeak,
strongly oriented, point against the surface. In the main peak, dipoles
are preferentially oriented toward the surface when localized on
top of O atoms. In the secondary layer, preferential polarization
is found in the whole plane against the surface.

\subsection{The Role of Electrostatics\label{sec:The-Role-of}}

\begin{figure}
\begin{centering}
\includegraphics[width=8.5cm]{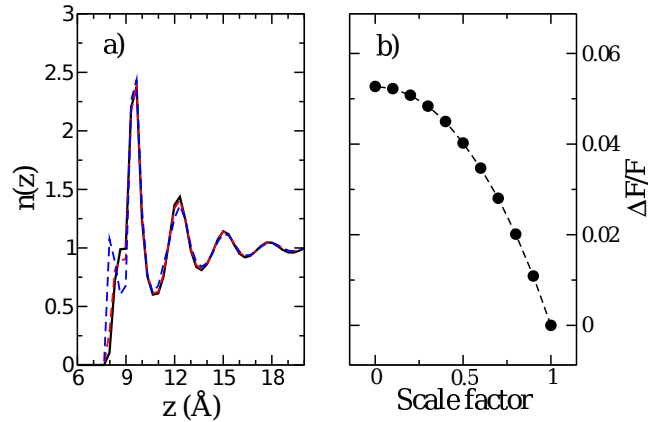}
\par\end{centering}
\caption{\label{fig:Averaged-density-along}a) Number density profile along
the $z$ axis with point charges of the clay atoms scaled by a factor
of 1 \textit{i.e.} completely turned on (full black line), 0.5 (dashed
red line), and 0 \textit{i.e.} turned off (blue dashed line). Only the
prepeak is affected. b) Relative change in solvation free energy as
a function of the scale factor. The relative changes stay below 5.5~\%.}
\end{figure}

The computational efficiency of the molecular density functional theory
 unlocks systematic
studies of the solvent structure and thermodynamics, such as the relative
roles of van~der~Waals and electrostatic interactions to the CLAYFF
force field. They are modeled respectively by Lennard-Jones interactions
and a point charge distribution reported in table~\ref{tab:Force-field-used}.
Figure~\ref{fig:Averaged-density-along} reports the number
density profile along the $z$ axis for modified systems where the
CLAYFF charges of the clay atoms have been scaled by a factor of 1,
0.5 and 0. It is observed that, surprisingly, only the shape of the prepeak is modified when the point
charges are turned off. This part of the density profile evolves from
a localized peak (with electrostatics on) to a shoulder of the main
peak when quenched. The rest of the number density profile is unchanged.
As might be expected, the polarization vanishes when charges are turned
off. In figure~\ref{fig:Averaged-density-along} is also plotted the relative
change in solvation free energy $F\left[\rho\right]$ for scale factors
between 0 (turned off) and 1 (completely turned on) of the charges
of the clay atoms. The solvation energy is affected by less than 6~\%
when charges are turned off. The contribution of the electrostatics
is thus small compared to the Lennard-Jones one, which is not an obvious result
even for a surface with a zero net charge.

\section{Conclusion\label{sec:Conclusion}}

We have shown that the molecular density functional theory in the
homogeneous reference fluid approximation (HRF-MDFT) is able to handle
the study of solvation properties of a complex clay surface of several
hundred atoms. The description of its structural and orientational
properties is quantitatively comparable to reference all-atom molecular
dynamics, while reducing the CPU-time by two orders of magnitude.
This means that the computation of these properties is now accessible
on a simple workstations within minutes. The HRF-MDFT calculations
allowed to accurately describe the subtle structure of the first molecular
layers at the surface, in particular the orientational properties. The
only noticeable difference lies in a localized zone of high density at
the center of hexagons, a defect due to a small overestimation of
the polarization close to the solute inherent to the method. Some
improvements in this respect are in progress.

The numerical efficiency of our approach made it possible to analyze the relative
contributions of the CLAYFF force field to the solvation properties.
The density profile and the solvation energy are found to be insensitive
to the electrostatic contribution. This may imply that the role of
electrostatics in charged clays may be reasonably reduced to their charged defects.
We think that this finding may be of high interest considering the large amount of work
dedicated to the derivation of accurate point charge distributions when building force fields
for such systems.

Finally, the next step of this work will be to consider the solvation of clay surfaces by a realistic model of water, at either a dipolar or multipolar level, and by ionic solutions.

\begin{acknowledgments}
The authors acknowledge financial support from the Agence Nationale de la Recherche under grant ANR-09-SYSC-012.
\end{acknowledgments}



\end{document}